\documentclass[9pt,twocolumn,twoside]{osajnl}
\journal{ol} 
\setboolean{shortarticle}{true}

\usepackage{listings}
\usepackage{caption}
\usepackage{subcaption}

\title{PyPhase - a Python package for X-ray phase imaging}

\author[1,*]{Max Langer}
\author[2]{Yuhe Zhang}
\author[4, 5]{Diogo Figueirinhas}
\author[6]{Jean-Baptiste Forien}
\author[1]{Claire Mouton}
\author[3, 5]{Rajmund Mokso}
\author[2]{Pablo Villanueva Perez}

\affil[1]{Univ Lyon, INSA‐Lyon, Université Claude Bernard Lyon 1, UJM-Saint Etienne, CNRS, Inserm, CREATIS UMR 5220, U1206, F‐69621, Villeurbanne, France}
\affil[2]{Division of Synchrotron Radiation Research and NanoLund, Department of Physics, Lund University, SE-221 00, Lund, Sweden}
\affil[3]{Division of Solid Mechanics, Faculty of Engineering, Lund University, SE-22100, Lund, Sweden}
\affil[4]{Division of Packaging Logistics, Faculty of Engineering, Lund University, SE-22100, Lund, Sweden}
\affil[5]{MAX~IV Laboratory, Lund University, SE-22100 Lund, Sweden}
\affil[6]{Lawrence Livermore National Laboratory, Livermore, CA, 94550, USA}

\affil[*]{Corresponding author: max.langer@creatis.insa-lyon.fr}

\begin{abstract}
X-ray propagation-based imaging techniques are well-established at synchrotron radiation and laboratory sources. However, most reconstruction algorithms for such image modalities, also known as phase retrieval algorithms, have been developed specifically for one instrument by and for experts, making the development and spreading of the use of such techniques difficult. Here, we present PyPhase, a free and open-source package for propagation-based near-field phase reconstructions, which is distributed under the CeCILL license. PyPhase implements some of the most popular phase-retrieval algorithms in a highly-modular framework supporting the deployment on large-scale computing facilities.  
This makes the integration, the development of new phase-retrieval algorithms, and the deployment on different computing infrastructures straight-forward. 
To demonstrate its capabilities and simplicity, we present its application to data acquired at synchrotron MAX~IV (Lund, Sweden). \\ 
\end{abstract}

\setboolean{displaycopyright}{false}

\begin{document}

\maketitle

%\section{Introduction}
The unique penetration power and short wavelength of X-rays make them an excellent probe to explore nature in a non-destructive manner down to the nanometer and the atomic scale. 
However, the high penetration power can lead to low contrast, especially for microscopic samples when exploiting the attenuation contrast. 
Propagation-based techniques can enhance the contrast and sensitivity by exploiting the phase contrast. X-ray propagation-based phase contrast imaging has seen a large increase in interest and development since its origins 25 years ago \cite{Snigirev1995, Wilkins1996}. These techniques have seen widespread use in imaging of soft tissues and fine variations in a dense matrix. With the use of phase retrieval algorithms together with computed tomography, the complex refractive index distribution in the imaged object can be reconstructed in 3D.

\begin{figure*}[!t]
    \centering\includegraphics[width=\linewidth]{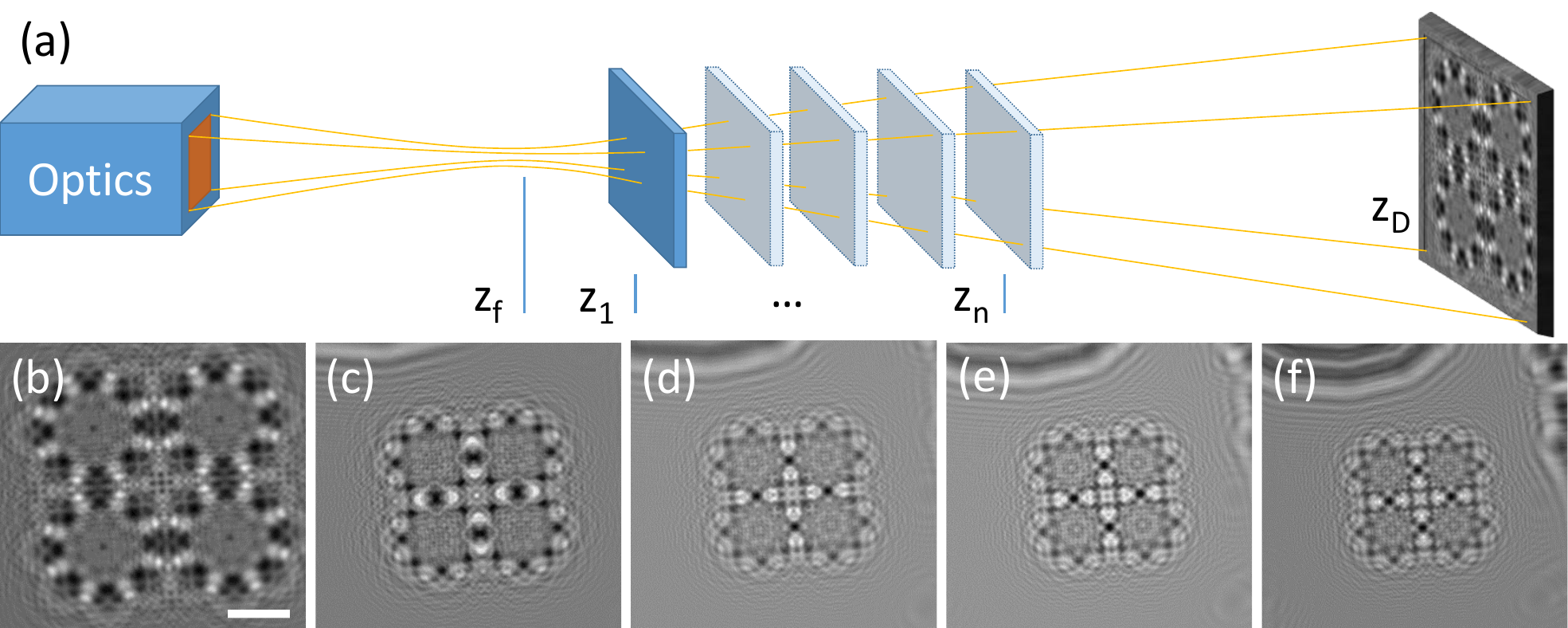}
    \caption{Phase-contrast imaging. (a) Schematic of the experimental setup for high-resolution phase-contrast imaging. An X-ray beam is focused using reflective (Kirkpatrick-Baez) optics. The sample is placed at different positions $z_n$ relative to the focus ($z_f$) and detector ($z_D$) positions for different amounts of magnification and consequently different effective propagation distances. (b)--(f) Phase-contrast images acquired at sample positions progressively further from the focus (and thus closer to the detector) showing the varying degree of magnification and phase contrast (acquired at NanoMAX, MAX~IV, Lund, Sweden). 
\label{fig:setup}}
\end{figure*}

Near-field propagation-based techniques record images that can be described as the intensity of the Fresnel transform of the exit wave as a function of energy, propagation distance, and length scale.
The first images recorded with these approaches had resolutions not better than the optical domain. 
More recently, the advent of X-ray optics has offered the opportunity to image at higher resolutions than conventional light microscopy~\cite{Mokso2007}.
Figure~\ref{fig:setup}(a) depicts a phase-contrast experimental setup using X-ray optics for high resolution imaging.
The images recorded by such experimental setups approach the holographic regime. In this regime, the direct interpretation of the images becomes more and more difficult with increasing resolution. 
Examples of holographic images from higher to lower magnification are shown in Figs.~\ref{fig:setup}(b--f).
Reconstructing the phase (and possibly the amplitude) of the exit wave using phase retrieval becomes a mandatory step for exploiting such images.

The phase-retrieval step has proved to be a persistent challenge, as no general algorithm currently exists. 
The practitioner is instead left to pick from a large range of algorithms that have been proposed in the literature. 
Initially, phase-retrieval algorithms were based on linearizations of the Fresnel integral, e.g., with respect to the exit wave itself~\cite{Gabor1948}. However, this solution is not applicable to all the imaging conditions and samples.

It is useful to classify propagation-based techniques into two regimes to understand their applicability and challenges.
These two regimes are divided with respect to the relative propagation distance.
Relatively short propagation distances with respect to the imaging length scale and wavelength yield what is sometimes called the \emph{edge-enhancement} regime. This translates to a Fresnel number $F=a^2/\lambda L$, with $a$ the size of the imaged feature, $\lambda$ the wavelength, and $L$ the propagation distance, corresponding to the smallest feature in the image that is $F\approx1$.
In this regime, phase contrast mainly enhances interfaces between materials in the sample with a visible diffraction fringe. 
Relatively long distances, corresponding to $F\ll 1$ but still within the validity of Fresnel diffraction, give rise to what is sometimes known as the \emph{holographic} regime, where the diffraction fringes dominate the contrast and are spread over a long distance in the object.

In the edge enhancement regime, algorithms based on assumptions on the imaged object, e.g., homogeneity of the constituent materials \cite{Paganin2002, Langer2010}, have seen widespread use. Such algorithms are simple and require only a single frame. 
These algorithms have been very successful for nearly homogeneous objects with fine variations in a dense matrix, but limitations of use remain for many types of samples due to violation of the homogeneity assumption.

More recently, algorithms taking into account the non-linearity of the problem have been proposed, which can be applied to the edge-enhancement and holographic regimes. Some take directly into account the specificity of the Fresnel framework \cite{Davidoiu2011}, whereas others borrow from the rich literature on reconstruction algorithms from coherent diffraction imaging and crystallography, based on projection onto sets type algorithms \cite{Gerchberg72ER,Fienup1980HIO,Elser2003}. 
Some of the latest developments on phase reconstruction seem to be pushing toward full 3D reconstruction \cite{Langer2012, Langer2013, Kostenko2013, Ruhlandt2014}, to be able to take into account constraints of the object in the phase retrieval procedure, instead of limiting to information from each projection in turn only. 

Although there is a plethora of literature regarding phase-reconstruction algorithms, access to those algorithms is relatively closed, with a few exceptions~\cite{Lohse2020, Weitkamp2011}. To address this limitation, we present a complete Python phase-retrieval package for X-ray phase-contrast imaging, named PyPhase. PyPhase is a fully open-source and modular package that relies solely on free and open-source tools. 

With PyPhase, we aim to provide:
\begin{itemize}
    \item A flexible phase-retrieval toolbox for expert users
    \item An interface, currently a command-line interface and in the future a graphical user interface, for non-expert users
    \item Tools for deployment on computer clusters and heterogeneous computing infrastructures
    \item Tools for implementation and development of phase retrieval algorithms 
    \item High level of modularity to facilitate the integration of different packages, e.g., registration, tomography, Fast Fourier Transform, reading and writing data, and visualization.
\end{itemize}

The package consists of a number of modules, namely: 

\begin{itemize}
    \item \emph{Phaseretrieval} contains functionality for phase retrieval algorithms. Table \ref{tab:edge_algorithms} lists the currently implemented algorithms.
    
    \item \emph{Propagator} contains functionality for the propagation of a wave-field and generation of intensities. The propagators are typically used in iterative phase-retrieval algorithms. Currently, a Fresnel operator is implemented, as well as linearized versions based on transport of intensity ~\cite{ReedTeague1983TIE} and contrast transfer equations~\cite{Guigay1977CTF}.
    
    \item \emph{Tomography} contains functionality for tomographic operations: forward projection, back-projection, as well as 3D image processing operators. The main part of this module wraps other codes for tomographic reconstruction, currently PyHST~\cite{Mirone2014} and  TomoPy~\cite{Gursoy2014}.
    
    \item \emph{Dataset} contains functionality to read and write images in different formats and from different data sources. Current data sources are ESRF style EDF data sets, NanoMAX style HDF5 data sets, and TOMCAT style TIF data sets. Compatibility with other data sources is planned, e.g., by interfacing DXChange \cite{DeCarlo2014}. 

    \item \emph{Utilities} contains supporting functionality, such as image registration and visualisation. Registration is currently implemented as a wrapper of Elastix~\cite{Klein2010, Shamonin2014} via pyElastix~\cite{PyElastix}. 

    \item \emph{Parallelizer} contains functionality for parallelization. This is implemented as decorators decorating functions that take a range of projections as input. Currently, the supported infrastructures are OAR \cite{OAR}, SLURM \cite{Slurm}, and serial processing. 
    Switching between infrastructures requires the modification of one line in the configuration file. 
\end{itemize}

Several phase retrieval algorithms are currently implemented in the package. On the one hand, some of the algorithms are appropriate for edge-enhancement regime due to model limitations on relative propagation distance or the necessity of a contact-plane image. On the other hand, appropriate algorithms for the holographic regime are provided, e.g., algorithms where the contact plane radiograph is not necessary, capable of reconstructing the attenuation and improving the high-frequency content (necessary for high-resolution imaging). The algorithms currently implemented in pyPhase, suitable for both edge-enhancement and holographic regimes, are listed in Table~\ref{tab:holo_algorithms}.

\begin{table}[!tbp]
    \caption{Phase retrieval algorithms available in PyPhase with their main applicability domain.}
    \centering
    \begin{tabular}{l l l} 
        \hline
        Edge enhancement regime & Class name & Source \\
        \hline\hline
        TIE, weak object (WTIE) & WTIE & \cite{Bronnikov2002} \\ 
        TIE, homogeneous objects & TIE-HOM & \cite{Paganin2002} \\
        Mixed approach, homogeneous & MixedApproach & \cite{Langer2010} \\
        Mixed approach, multi-material & MixedApproach & \cite{Langer2012} \\
        Mixed approach, heterogeneous & MixedApproach & \cite{Langer2013} \\
        \hline
        \hline
        Holographic regime & Class name & Source \\ [0.5ex] 
        \hline\hline
        Contrast Transfer Function (CTF) & CTF & \cite{Cloetens1999} \\ 
        CTF, pure-phase object & CTFPurePhase & \cite{Cloetens1996} \\
        %ADMM-CTF & ADMMCTF & %\cite{Villanueva-Perez2017} \\
        Gradient descent & GD & \cite{Langer2012b} \\
        Hybrid Input-Output (HIO) & HIO\_ER & \cite{Fienup1978} \\
        Error Reduction (ER) & HIO\_ER & \cite{Gerchberg72ER} \\ 
        \hline
    \end{tabular}
    \label{tab:edge_algorithms}
    \label{tab:holo_algorithms}
    %\vspace{-4mm}
\end{table}

There are many more algorithms presented in the literature than currently covered by PyPhase. 
The aim, however, is to progressively include as many algorithms as possible. 
This would permit a fair evaluation of algorithms in practice.
Furthermore, it would help the practitioner choose the most suitable algorithm, since different algorithms work well in different circumstances. 

The package contains classes representing the building blocks of phase retrieval algorithms. This modularity makes it simple to swap between different implementations, e.g., one optimized for GPU or CPU processing, without substantial changes to the code. The most basic operation for phase retrieval is arguably the FFT. For flexibility, the package contains a wrapper for the FFT functions (FFT, IFFT, fftshift, ifftshift, frequency variables). In the current version, the numpy FFT is used. Interfaces to other FFT packages, such as FFTW, can then be easily implemented and used with minimal modification of the code.

\begin{figure*}[tbp!]
\centering\includegraphics[width=\linewidth]{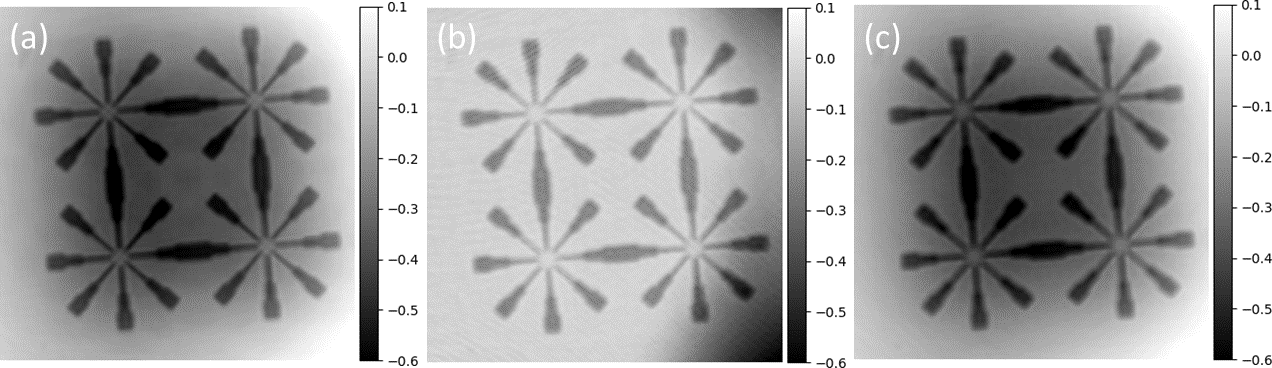}
    \caption{Phase-retrieved results obtained with PyPhase using (a) the CTF pure phase algorithm, (b) average of 5 iterations of successively 45 iterations of HIO, followed by 5 iterations of ER, separately on each acquired image, and (c) 20 iterations of gradient descent initialised with the CTF pure phase reconstruction. The colorbar represents the phase-reconstructed values in radians.
    \label{fig:results}}
\end{figure*}

Iterative phase retrieval algorithms require the implementation of propagators. An implementation of the Fresnel transform as well as a linearized propagator based on the CTF are provided. The modular implementation makes it straight-forward to interface with other codes and algorithms.

For the implementation of 3D phase-retrieval algorithms, it is necessary to implement the Radon transform and its inverse transformation. 
These tomographic operations are provided as wrappers of specialized tomographic-reconstruction packages. 
Currently, Tomopy and PyHST2 packages can be used. 
However, the modularity of PyPhase makes interfacing other packages straight-forward. 

A limiting factor for the practical application of phase retrieval in tomography
is efficient data handling. 
We implement this as a dataset class, with subclasses that interface different sources, which provides the basic reading and writing of images in a tomographic or near-field imaging setup.
Currently, classes for ESRF style EDF data, Tomcat style TIF data, and NanoMAX style HDF5 data are implemented. 
The modularity of the code makes it simple to add new data sources, e.g., by interfacing DXchange \cite{DeCarlo2014}. 
This facilitates the treatment of data from different sources using the same reconstruction parameters. 
% Repetition? The other mention should be purely module structure

Another limiting factor for phase retrieval in tomography is the computation time. 
Parallelization efforts are often centered on GPU processing, neglecting the large availability of computing clusters and heterogeneous computing architectures. 
Since the projections in a tomographic data set are usually considered independent, parallelization over the projections is "embarrassingly parallel". 
To leverage this, a function decorator \emph{@Parallelize} is provided, which can be applied to all functions in the package that take a range of projections as an argument. 
It allows to automatically split the implicit \emph{for} loop into appropriate chunks and farm them out on the desired number of CPUs.
The infrastructure used can be switched with a centralized parameter in a configuration file, thus requiring no modification of the code to change from serial to parallel computing. 
Currently, supported infrastructures are serial, for workstation use, SLURM and OAR resource and task managers, as well as serial execution. 
The modular implementation makes it straight-forward to include other infrastructures.

A certain number of auxiliary functionality important for phase retrieval is implemented. 
Most important is image registration, which is used to align the images at one projection angle. 
Phase-contrast images at different propagation distances and magnifications do not contain identical contrast. 
This effect precludes simple correlation-based approaches in all but the most trivial cases. 
For this case, registration algorithms based on mutual information~\cite{Weber2018} seem to be more appropriate.
Image registration is implemented as a class \emph{Registrator}. 
Currently, this is implemented as a wrapper for the Elastix software through the PyElastix interface. 
Elastix provides all the necessary functionality, but it might not be available on all platforms. 
In that case, the modularity makes it straight-forward to integrate other image registration codes.

A simple interface for the visualization of the different images is provided. 
This is to simplify the access to images, e.g., corrected and non-corrected projections, for verification purposes.
Furthermore, it enables plotting curves related to the reconstruction process, e.g., the image alignment and regularisation parameter choice. 
The current implementation relies on the Matplotlib package, but more advanced visualization could be used exploiting PyPhase modularity.

The number of dependencies is kept as low as possible. All arrays and matrix-vector operations are implemented using NumPy \cite{Numpy}. 
Other dependencies are currently i) Elastix for image registration, since no adequate pure Python implementation is available, and ii) PyHST2 or TomoPy for tomographic operations.

The implementation choices yields extremely compact code. For example, the code for the reconstruction of the image in Fig.~\ref{fig:results} (a) using the CTF Pure Phase algorithm is:
\begin{minipage}{\linewidth}
%\vspace{2mm}
    \begin{lstlisting}[language=Python]
    import pyphase
    data = dataset.Nanomax('star')
    data.AlignProjections()
    retriever = phaseretrieval.CTF(data)
    retriever.ReconstructProjections()
    \end{lstlisting}
\end{minipage}
Reconstructions obtained with PyPhase using a similar code for three different algorithms are depicted in Fig.~\ref{fig:results}. A \href{https://pyphase.readthedocs.io}{documentation and API} is available \cite{PyphaseAPI}.

To conclude, we have presented an open-source and modular Python package for phase retrieval, christened PyPhase. 
PyPhase has the potential to lower the entry barrier to phase retrieval in the Fresnel regime and thus make the technique more accessible to non-expert users. 
The package covers the most popular phase-retrieval algorithms presented in the literature. 
Furthermore, it provides tools and an interface that facilitate the implementation and deployment of other phase-retrieval algorithms.
To further facilitate its usability by non-expert users, we provide a command-line interface and, in the future, a graphical user interface.
Contributions are welcomed: contact the authors for more information or go directly to the project \href{https://gitlab.in2p3.fr/mlanger/pyPhase}{repository} \cite{PyphaseRepository}.
\section*{Acknowledgments}
We are grateful to S. Kalbfleish and M. Kahnt for their support acquiring and implementing propagation-based phase contrast imaging at NanoMAX, MAX~IV, Lund, Sweden. Jean-Baptiste Forien acknowledges support by the US Dept. of Energy (contract No. DE-AC52-07NA27344) and by the Office of Laboratory Directed Research and Development (LDRD), tracking numbers 19-ERD-022. 
\bibliography{pyphase}
\end{document}